\documentclass[preprint]{aastex}
\begin{document}
\title{A Far Ultraviolet Study of the Hot White 
Dwarf in the Dwarf Nova WW Ceti
\altaffilmark{1}}

\author{Patrick Godon\altaffilmark{2}}
\affil{Department of Astronomy and Astrophysics
Villanova University,
Villanova, PA 19085,
patrick.godon@villanova.edu}

\author{Laura Seward}
\affil{Dept. of Astronomy,
Florida Institute of Technology, Melbourne, FL,
lseward@fit.edu}

\author{Edward M. Sion}
\affil{Dept. of Astronomy \& Astrophysics,
Villanova University,
Villanova, PA 19085,
edward.sion@villanova.edu}

\author{Paula Szkody}
\affil{Department of Astronomy,
University of Washington,
Seattle, WA 98195,
szkody@astro.washington.edu}

\altaffiltext{1}
{Based on observations made with the 
NASA-CNES-CSA Far Ultraviolet Spectroscopic
Explorer. FUSE is operated for NASA by the Johns Hopkins University under
NASA contract NAS5-32985} 
\altaffiltext{2}
{Visiting at the Space Telescope Science Institute, Baltimore, MD 21218,
godon@stsci.edu}

\begin{abstract}

We present a synthetic spectral analysis of 
{\it{IUE}} archival and {\it{FUSE}} 
FUV spectra of the peculiar dwarf nova WW Ceti. 
During the
quiescence of WW Ceti, a white dwarf with 
$T_{wd}\sim 26,000$K $\pm 1000$K 
can account for the FUV flux and yields the proper distance. However,
the best agreement with the
observations is provided by a two-temperature white dwarf model with a
cooler white dwarf at $T_{wd} = 25,000$K providing 75\% of the FUV flux
and a hotter region (accretion belt or optically thick disk ring) with $T
= 40,000$K contributing 25\% of the flux for the proper distance. We find
from the {\it{FUSE}} spectrum that the white dwarf is rotating with 
$V \sin{i} = 600 \pm 100 $km$~$s$^{-1}$. 
Our temperature results provide
an additional data point in the distribution of WD $T_{wd}$ versus
orbital period above the CV period gap where few WD $T_{wd}$s are
available.

\end{abstract}


\section{Introduction}

Cataclysmic variables (CVs) are short-period, semi-detached binary systems
consisting of an accreting white dwarf (WD) primary star and a low-mass
main-sequence secondary star as the Roche lobe-filling mass donor
\citep{war95}. 
The binary orbital period in these systems ranges from about $\approx 1$h 
to a few days. However, there is a gap in orbital period
between 2 and 3 hours where almost no systems are found. 
It is not known 
whether the systems are evolving from a longer period to a shorter
period (across the gap) or whether the systems above the gap are all together
different from the systems below the gap.
Dwarf novae (DN) are a subset of CVs that undergo regular
eruptions called outbursts which
last for days to weeks, separated by intervals of quiescence
lasting weeks to months.
The now widely accepted interpretation of the quiescence/outburst
cycle is that of the disk instability model (DIM, \citet{can98}).  
It is assumed that during the quiescent
phase the matter in the disk is cold and neutral and the
disk is optically thin because of its low density, while 
during outburst, as the mass accretion rate increases, the matter in the 
disk is ionized and becomes optically thick. The basic principle 
of the DIM theory depends 
heavily on the unknown viscosity parameter $\alpha$ and 
on the mass accretion rate during the different phases. 
The mass accretion rate is usually taken from \citet{pat84}, 
which is however only a first order estimate. 
In the last decade, the  
mass accretion rate of many systems has 
been deduced more accurately at given epochs
of outburst or quiescence using spectral fitting
techniques. The accretion rate is usually a function of time
(especially during the outburst itself) and consequently  
it is difficult to assess its time-averaged value accurately. 

Recent advances in theory \citep{tow04}  have shown that
the average mass accretion rate of an accreting WD in DNe can be
deduced if one knows the mass of the accreting WD and
its effective surface temperature during quiescence, therefore
providing an additional and independent way to assess $\dot{M}$.  
Consequently, in order to put more constraints 
on the theories we need to known the
properties (mainly the temperature and mass of the WD) 
of these systems above as well as below the period gap. 
There is, however, a critical shortage 
in knowledge of the WD properties (effective temperature $T_{wd}$, 
gravity $\log{g}$, rotational velocity 
$V \sin{i}$, chemical abundances, accretion belts?) 
in dwarf novae above the period gap. 
Thus, detailed comparisons of accreting WDs above and 
below the gap cannot be made. 

For systems below the gap, with 
orbital periods near the period minimum, the distribution of temperatures 
are centered at $\sim$15,000K with only a narrow range seen at present. 
This distribution appears to manifest the effect of long term 
compressional heating at a time averaged accretion rate of $2\times 
10^{-11} M_{\odot}$yr$^{-1}$  
(Townsley and Bildsten 2002; Sion et al. 2003). 
It appears that WD $T_{wd}$'s for systems above the gap are higher than WD 
temperatures in systems below the gap, due to the systems above the gap 
having larger disks (with higher mass transfer rates) and more massive 
secondaries.  Some disks may remain optically thick even during quiescence 
so that the WDs are heated to a greater extent than systems below the gap. 
It is not yet known whether the WDs in systems above the gap are
rotating more slowly than WDs in systems below the gap where
presumably the CVs are older with a longer history of angular momentum
transfer via disk accretion. Thus far the only dwarf novae above the gap 
whose white dwarfs and disks/boundary layers 
have been analyzed with {\it{FUSE}}, {\it{IUE}} and {\it{HST}} have been 
Z Cam \citep{har05}, RX And \citep{sio01,sep02},  
U Gem \citep{sio98,lon99,fro01}, SS Aur \citep{sio04a},   
EY Cyg \citep{sio04b}, and RU Peg \citep{sio02,sio04a};    
a total of 2 Z Cam systems and 4 U Gem systems. 

WW Ceti is a system above the gap 
whose classification is uncertain. Although classified
as a U Geminorum-type system, others suggest a Z Camelopardis
classification. Z Cam systems are characterized by prolonged states of
intermediate brightness between quiescence and outburst called
standstills. Others
propose that WW Ceti forms a link between Z Cam and VY Sculptoris
nova-like variables \citep{rin96,war87}.  The latter are systems that drop
unpredictably into very low brightness states but spend most of their time
in outburst. Without conclusive
evidence of standstills however, we will tentatively classify WW Ceti as a
U Gem-type DN.

In this paper, we report an analysis of International Ultraviolet Explorer
({\it{IUE}}) archival spectra and Far Ultraviolet Spectroscopic Explorer
({\it{FUSE}}) spectra of the dwarf nova WW Ceti during quiescence. Our
analysis utilizes newly available accretion disk models, photosphere
models, and models combining white dwarfs and accretion disks and
accretion belts. The accretion disk models are taken from the grid of models  
of \citet{wad98}, which were recently updated using the lastest version
of the stellar/accretion disk synthetic spectral codes (see section 3). 
Our objectives are
to identify the source(s) of the FUV radiation during quiescence, derive
the properties of the WD (if detected) and the quiescent accretion disk, and
characterize the hot components in the system.

\subsection{WW Ceti System Parameters}

In table 1 we list the system parameters we have adopted: (1) CV subtype,
(2) orbital period in days, (3) orbital inclination in degrees, (4)
spectral type of the secondary, (5) mass of the primary in solar masses,
(6) mass of the secondary in solar masses, (7) apparent magnitude in
outburst, and (8) apparent magnitude in quiescence. The references are
listed below the table.

The orbital period (4.22 hours) is well above the period gap, where few WD
properties are currently known. Systems above the gap tend to have
somewhat earlier-type secondaries, higher accretion rates, and larger
accretion disks. 
Dwarf novae, unlike other CVs, offer a fairly reliable estimate of
their distances via the absolute magnitude at maximum versus orbital
period relation for dwarf novae found by \citet{war95}. This relationship
is consistent with theory \citep{can98}. 
For WW Ceti, this
relation yields a distance of 186 pc. This is midrange of $90 - 300$ pc as
derived by \citet{you81} using near-infrared CCD spectra of the cool
companion, slightly higher than the estimate of 130 pc \citep{pat84} and
close to the range of $121 - 171$ pc using K magnitudes \citep{spr96}.  
To remain consistent with the parameters used in \citet{win03}, we adopted
186 pc as the distance to WW Ceti.

\section{The Observations}

\subsection{The {\it{FUSE}} Observations}

{\it{FUSE}} is a low-earth orbit satellite, launched in June 1999. Its optical
system consists of four optical telescopes (mirrors), each separately
connected to a different Rowland spectrograph. The four diffraction
gratings
of the four Rowland spectrographs produce four independent spectra on
two photon
counting area detectors. Two mirrors and two gratings are coated
with SiC to provide wavelength coverage below 1020 \AA, while the other
two mirrors and gratings are coated with Al and LiF overcoat.  The
Al+LiF coating provides about twice the reflectivity of SiC at
wavelengths $>$1050 \AA, and very little reflectivity below 1020 \AA\
(hereafter the SiC1, SiC2, LiF1 and LiF2 channels).  

A time tag {\it{FUSE}} spectrum (D1450401) of WW Ceti was obtained
starting on 2003, July 26, 05:30:03 with a total raw exposure time 
of 16,291s
(9 individual spacecraft orbits) through the 30"x30" LWRS
Large Square Aperture. The spectrum was centered at 1059.73\AA. 
From the {\it{AAVSO}} (American Association of Variable Stars Observers)
data, we found that the system was in optical quiescence
($M_V \approx 14.5-15.0$ for at least 10 days) preceding and
during the observation,
and 25 days after an outburst ($M_V \approx 12.0$)
and about 15 days after a single data point brightness peak
($M_V \approx 11.0$).  The data were 
processed with CalFUSE version 2.4 totaling 14,817s of good exposure time. 
In this version, event bursts are automatically taken care of. 
Event bursts are short periods during an exposure when high count 
rates are registered on one of more detectors. The bursts exhibit  
a complex pattern on the detector, their cause, however, is yet unknown 
(it has been confirmed that they are not detector effects). 
WW Cet, with a flux of a few $ 10^{-14}$ergs$~$s$^{-1}$cm$^{-2}$\AA$^{-1}$,
is actually a relatively weak source. 
We used the same procedure as in our previous {\it{FUSE}} analysis
(e.g. RU Peg \&  SS Aur, \citet{sio04a})
to process the {\it{FUSE}} data.  

During,
the observations, Fine Error Sensor A, which images the LiF1 aperture
was used to guide the telescope. The spectral regions covered by the
spectral channels overlap, and these overlap regions are then used to
renormalize the spectra in the SiC1, LiF2, and SiC2 channels to the flux in
the LiF1 channel. We then produced a final spectrum that covers almost the
full {\it{FUSE}} 
wavelength range $905-1182$ \AA. The low sensitivity portions of
each channel were discarded.
In most channels there exists a narrow dark stripe of decreased flux
in the spectra running in the dispersion direction. This stripe has been
affectionately known as the ``worm'' and it can attenuates as much as
50\% of the incident light in the affected portions of the
spectrum. The ``worm'' has been observed to 
move as much as 2000 pixels during a 
single orbit in which the target was stationary. The ``worm'' appears to be
present in every exposure and, at this time, there is no explanation for it. 
Because of the temporal changes in the strength and position of the 
``worm'',
CALFUSE cannot correct target fluxes for its presence. 
Here we took particular care to discard the portion of the spectrum 
where the so-called {\it{worm}} 'crawls'.
The LiF1 channel was affected by the worm longward of $\approx 1100$ \AA\ . 
Because of this the $1182 - 1187$ \AA\ region was lost, as it is
covered only by the LiF1 channel. 
We combined the individual exposures and channels to create a
time-averaged spectrum with a linear, $0.1$ \AA\ dispersion, weighting
the flux in each output datum by the exposure time and sensitivity of the
input exposure and channel of origin. 

The {\it{FUSE}} spectrum of WW Cet (see Figure 1) in the longer
wavelengths ($\lambda > 1050$\AA) is very similar to the {\it{FUSE}}
spectra of DNe in quiescence in which the WD is exposed.  Here, that
portion of the spectrum reveals the familiar absorption features of
C{\small{III}} (around 1175\AA)  and Si{\small{III}} (around 1110\AA,
1140-1145\AA, and 1160\AA).  However, at shorter wavelengths ($\lambda
< 1050$\AA) there are very broad emission lines that are usually found in
Nova-like systems (e.g. such as AE Aqr, V347 Pup, DW UMa, see the MAST
{\it{FUSE}} archives) and are not usually present in low-inclination
DN systems in quiescence.  
The most prominent emission feature is the O{\small{VI}} doublet 
with laboratory wavelengths 1031.9261\AA\ \& 1037.6167\AA .  
The feature is best fitted with a double Gaussian shifted by $+1.2$\AA\
(see Figure 2, upper panel). A double Gaussian at the
rest frame (Figure 2, lower panel) does not fit the data as well
and a single Gaussian disagrees even more with the O{\small{VI}}
emission feature.  
On top of the O{\small{VI}} doublet is a a C{\small{II}}
interstellar absorption at the expected rest-frame wavelength 1036.3\AA.
The C{\small{III}} (977.02\AA) broad emission line seems also to
have a C{\small{III}} interstellar absorption superposed to it. 
The two N{\small{III}} and the Si{\small{III}} emission lines
(between 980\AA\ and 1000\AA)  are also marked, but they are contaminated
with some air glow emission (O{\small{I}}) and are less prominent than the
C{\small{III}} and O{\small{VI}} emission features. For a more complete
comparison we have also marked the positions of the 
N{\small{IV}} emission lines about 925\AA\ , which form a broad
``hump'' (and could possibly also include S{\small{VI}}).  The
{\it{FUSE}} spectrum of WW Ceti also exhibits some sharper absorption lines,
probably of interstellar origin, the most prominent ones are 
N{\small{I}} (around 1134\AA), N{\small{II}} (around 1084-1085\AA), 
C{\small{II}} (1036.3\AA) and C{\small{III}} (977\AA).

\subsection{The {\it{IUE}} Archival Observations} 

The {\it{IUE}} archive contains two observations of WW Ceti with the
short-wavelength prime (SWP) camera through the large aperture at low
dispersion with a resolution of 5\AA , covering the wavelength range 1170
to 2000\AA. The brightness state of the system at the time the {\it{IUE}}
spectra were taken was determined by the observed flux levels and
{\it{IUE}} fine error sensor visual magnitudes of the system as
confirmed by light-curve data from the American Association of Variable
Star Observers (AAVSO). One of the spectra, SWP24866, was obtained during
quiescence, its  observing log is shown in Table 2.
In the present work we concentrate only on the quiescent phase  
of the system. 

The {\it{IUE}} NEWSIPS spectrum was flux-calibration corrected using the 
Massa-Fitzpatrick corrections \citep{mas00}. 
The spectrum was not corrected for
reddening, as E(B-V) = 0 for WW Ceti 
\citep{lad91}. 
The {\it{IUE}} spectrum also exhibits some broad emission lines. The most
prominent ones are C{\small{IV}} (1548/50\AA), He{\small{II}}
(1640\AA\ Balmer $\alpha$), O{\small{III}} (1658-1666\AA) and the
strong feature around 1400 \AA\ is either due to either O{\small{IV}}
or to Si{\small{IV}} or to both. The spectrum is very similar
to some NL systems (IP).
With their broad emission lines, both the {\it{FUSE}} and {\it{IUE}} spectra
show evidence of hot gas.

A comparison of the {\it{IUE}} spectrum in quiescence and the {\it{FUSE}}
spectrum in the wavelength overlap region reveals that the flux levels
match thus enabling us to carry out model fits over a substantially
broader wavelength range. We prepared the combined {\it{FUSE}} and
{\it{IUE}} spectra for fitting by masking regions containing emission
lines and artifacts. The following wavelengths were masked: 
$ 949 - 951, 972 - 995, 1025 - 1043, 1168 - 1170, 
1190 - 1240, 1387 - 1412, 1517 - 1565, 1620-1680, 1840 - 1878$\AA.
These regions of the spectrum were not included in the fitting.

\section{Synthetic Spectral Fitting}

We used model accretion disks from the optically thick disk model grid of
\citet{wad98} and created model spectra for
high-gravity stellar atmospheres using TLUSTY200 \citep{hub88}, SYNSPEC48
and ROTIN4 \citep{hub95}.  Using IUEFIT, a $\chi^2$ minimization routine,
we computed $\chi^{2}_{\nu}$ and scale factor values for each model fit.  
$\chi^2_{\nu}$ is known as the ``reduced'' chi-squared, namely $\chi^2$
per number of degrees of freedom. The number of degrees of freedom $\nu$ 
is the number of bins $N_{bins}$ 
for which observed data is taken into account when calculating
$\chi^2$ minus the number of freely varying parameters $N_p$ used
to calculate the model, namely $\nu = N_{bins}-N_p$, 
see e.g. \citet{numrec}. 
Here we take the number of bins in the wavelength axis excluding 
the masked regions. 
For the {\it{IUE}} data binned at 1.68 \AA\ 
$N_{bins}=370$, for {\it{FUSE}} binned at 0.1 \AA\ $N_{bins}=2293$
(and decreases accordingly for binning of 0.2 and 0.5 \AA\ ) and for
the combined {\it{IUE-FUSE}} spectrum $N_{bins}=817$ (as the 
{\it{FUSE}} data there are binned at 0.5 \AA\ ).  

The range of disk model parameters are the following: WD mass (in solar
masses)values of 0.55, 0.80, 1.03, and 1.21; orbital inclination $i$ (in
degrees) of 18, 41, 60, 75, and 81. The accretion rate ranges from
$10^{-10.5} M_{\odot}$yr$^{-1}$ to $10^{-8.5} M_{\odot}$yr$^{-1}$, varying in
increments of 0.5 in $\log{\dot{M}}$.
Therefore, for the disk models one has
$N_p=3$. For the photosphere models, the effective
temperature values range from 15,000 K to 35,000K in increments of 1,000K.
We chose values of $\log{g}$ ranging between 8.0 and 8.6
for consistency with the observed mass. We also varied the
stellar rotational velocity $V \sin{i}$ from $100$km$~$s$^{-1}$
to $1000$km$~$s$^{-1}$ in steps of $100$km$~$s$^{-1}$. In order to
try and fit the absorption features of the spectrum, we also vary
the chemical abundances of C, N and Si.
The number of freely varying
parameters for the stellar atmosphere models can therefore be as
high as $N-p=6$.  
For combined accretion disk and stellar WD atmosphere models, the disk 
flux is divided by 100 to normalize it at 1000pc to match the WD flux, 
therefore giving explicitly the relative flux contributions of each
component. Then both fluxes are added for comparison with the observed flux. 
For the combined models this leads to $N_p=8$, since for any WD 
mass there is a corresponding radius, or equivalently one single value of
$\log{g}$ (e.g. see the mass radius relation
from \citet{ham61} or see \citet{woo90} for different
composition and non-zero temperatures WDs).
While we use a $\chi^2$ minimization technique, we do not 
``blindly'' select the least $\chi^2$ models, but we examine the models 
that best fit some of the features such as absorption
lines (see the fit to the {\it{FUSE}} spectrum alone) 
and, when possible, the slope of the wings of the broad Lyman
absorption features. We also select the models that are in agreement
with the known distance of the system.  

\subsection{Model Fits to the {\it{IUE}} Archival Spectra }

It is useful to explore the results of analyzing the {\it{FUSE}} and
{\it{IUE}} spectra separately versus when they are combined. We started
with the {\it{IUE}} archival spectrum SWP24866, taken in quiescence.
Applying the grid of disk models to the {\it{IUE}} data, the best-fitting
disk-only model had a WD mass of $M_{1} = 0.80 M_{\odot}$, an orbital
inclination of 60 degrees, and an accretion rate of $10^{-9.5}
\dot{M}$yr$^{-1}$. This model gave a $\chi^{2}_{\nu}$ 
value of 1.870 and a scale
factor of 0.063, which yielded a distance of 395 pc. This accretion rate
is rather large to be associated with quiescence and the distance from
the scale factor is far larger than the adopted distance of 186 pc.
Therefore we reject this result.

For single temperature white dwarf models, the best-fitting model 
photosphere had an effective temperature of 23,000 K, $\log{g}$= 8.0, 
giving a $\chi^2_{\nu}$ value of 2.97 and a scale factor of 0.004717, 
which yielded a 
distance of 199 pc. This distance is in excellent agreement with our
adopted distance. This result suggests that a white dwarf with a T$_{wd}
= 23,000$K could alone account for the FUV flux without an accretion disk.

\subsection{Synthetic Spectral Analysis of the {\it{FUSE}} Spectrum Alone} 

We first tried an accretion disk alone, but the fit was extremely poor.  
We then tried to fit the {\it{FUSE}} spectrum with a WD alone synthetic
spectrum. The best fit model was a $27,000 \pm 1,000$K WD, with
$\log{g}=8.3$, a rotation rate of $600 \pm 100$km$~$s$^{-1}$, a distance
of 212pc and $\chi^2_{\nu}=1.58$. 
The abundance were C=0.1 x solar, N=2 x solar and Si=0.3 x solar.  
A solar abundances model had a $\chi^2_{\nu}=1.629$, not statistically
different from the non-solar model, however some of the features
of the observed {\it{FUSE}} spectrum were better fitted with non-solar 
abundances. The results are presented in Figure 3.  

In particular, fitting the following features were the main reason for
adopting the non-solar abundances model as the best fit. \\  
\indent 
(i) In the solar abundances model there is a broad absorption feature
around 1065 which is not present in the observed {\it{FUSE}} spectrum.
When decreasing C this absorption feature disappeared and we obtained
a better fit. This absorption feature is due to \\ 
C{\small{II}} 
1063.28, 1063.31, 1065.89, 1065.92, 1066.13 \AA\ . \\  
\indent 
(ii) There is a deep absorption feature in the {\it{FUSE}} data around
1084-1085 A, although this might be due to interstellar absorption
we found that increasing N by a factor 2 solar gives a better fit.
This absorption feature is due to \\ 
N{\small{II}} 
1083.99, 1084.56, 1084.58, 1085.53, 1085.55, 1085.70 \AA\ . \\  
\indent 
(iii) Around 1108-1114, 1123, 1128, 1140-1146 and around 1160 \AA\ there
are many absorption features which can be matched only by decreasing
C and Si abundances.
These are : \\ 
Si{\small{III}}: 
1108.36, 1109.94, 1113.17, 1113.20, 1113.23 \AA\ \\ 
blend: C{\small{III}} 1125.6 + C{\small{II}} 1127. + 
Si{\small{IV}} 1128.3 \AA\ + C{\small{I}}  \\ 
blend: C{\small{I}}+C{\small{II}}+C{\small{III}} 1138-1141 \AA\  \\ 
blend: Si{\small{III}} 1140.5-1145.7 \AA\ .   \\ 

While it is obvious that there must be some 
C{\small{III}} broad emission around 1175 \AA\,
there is also absorption present. 
Since we do not know a priori how much emission
and absorption there is, we cannot model this region. 

Decreasing C and Si further than 0.1 and 0.3 (respectively) 
led to improvement of the fit in
some parts of the spectrum but the fit also deteriorated in 
other regions. It is also important to note that the
depth of the broad absorption features which actually form part of 
the continuum (e.g. as for Si) depends not only on the abundances
but also on the rotational velocity. Increasing the rotational
velocity has the same effect as reducing the abundances, as the
depth of the absorption features decreases. Overall, the
C{\small{II}} (around 1065 \AA\ ), the Si{\small{III}} 
(around 1110 \AA\ ) and the Si-C blend (in the range 1138-1146 \AA\ ) 
were the main absorption features  that drove both the Si and the C
to sub-solar values. As these abundances 
were reduced, the fit of the theoretical
spectrum also slightly improved in other regions (such as around
1055 \AA\ and 1080 \AA\ ).  
 
We feel confident that the carbon and silicon WD photospheric 
abundances are
actually well represented by the non-solar best fit models, as these
were derived from fitting these regions of the spectrum that
were not contaminated by broad emission features and/or sharp
interstellar absorption features. For Nitrogen,
the results are inconclusive as the Nitrogen features were contaminated
by sharp absorptions possibly originating from the ISM or from the
immediate surroundings of the system. 

In Figure 4 we show the residual
emission, namely, we subtracted the theoretical spectrum from the
observed one. The apparent emission features in the residual spectrum are
the O{\small{VI}} doublet (with an interstellar C{\small{II}} absorption),
the two C{\small{III}} lines 
(977 \AA\ and 1175 \AA\ ) and possibly also
N{\small{II}}, N{\small{III}}, N{\small{IV}} and Si{\small{III}} as marked
in Figure 4. Such emission features are often seen in the {\it{FUSE}}
spectra of IP systems or other DNe systems and it implies that gas is
possibly being heated by shock (e.g. accretion column in magnetic
systems).

From these fits, it is evident that
a WD alone is not enough to reproduce all the features of the spectrum.
The bottom of Ly$\beta$ does not go to zero and indicates the
presence of an additional component, which may or may not be 
linked to the broad emission features. 
While it is clear that the longward wing of
the Ly$\beta$ cannot go to zero because of O{\small{VI}} emission, the
shortward  wing reaches 
$1 \times 10^{-14}$erg$~$s$^{-1}$cm$^{-2}$ \AA$^{-1}$.  
We tried a combination of a WD plus an accretion disk and, as in the case
of the {\it{IUE}} spectrum, we found that the best fit model had a
distance about twice as large as the adopted distance of 186pc.  
Therefore, we tried a combination of a hot, fast rotating belt with a WD.
The best combined (WD+belt) model consisted of a WD with a temperature of
25,000K, rotating at 600km$~$s$^{-1}$, with $\log{g}=8.3$ , and with the
following abundances: Carbon 0.1 x solar, Nitrogen 2 x solar and Silicon
0.5 x solar. The second component had a temperature of 40,000K, rotating
at 3,000km$~$s$^{-1}$ and had an emitting area of 3 percent of the total
WD surface. The WD contributes 75\% of the flux integrated over the
{\it{FUSE}} wavelength range, while the second component contributes the
remaining 25\%. The $\chi^2_{\nu}$ value was 1.38 with a distance of 190pc. 
This best fit is shown in Figure 5.

\subsection{Synthetic Spectral Analysis of Combined {\it{FUSE}} 
plus {\it{IUE}} Spectra}

We combined the {\it{FUSE}} and {\it{IUE}} quiescent 
spectra of WW Ceti 
and first tried single-temperature white dwarf models to the
combined spectra. The best-fitting model for $\log{g} = 8.3$ yielded the
following parameters:  $\chi^2_{\nu}=3.73$, d=189pc, T$_{wd} = 26,000\pm
1000$K, $V\sin{i} = 600 \pm 100$ km$~$s$^{-1}$, and abundances: Carbon
0.1 x solar, Nitrogen 2 x solar and Silicon 0.5 x solar. The white dwarf
rotation ($V\sin{i}$) rate was determined from fitting the
WD model to the spectrum while paying careful attention to the 
line profiles in the {\it{FUSE}} portion of the combined spectrum.
Namely, we did not carry out separate fits to individual lines but
rather tried to fit the lines and continuum in the same fit. The
best-fitting single temperature WD fit to the combined 
{\it{FUSE}} plus {\it{IUE}} spectrum is displayed in figure 6.

Second, we tried models of optically thick accretion disks alone.  Since
the previous individual disk fits to the {\it{FUSE}} and {\it{IUE}}
spectra alone were very poor, we did not expect the disk fit to the
combined spectrum to be better. For this exercise, we fixed the disk
inclination at the value published for WW Ceti. We also fixed the white
dwarf mass at 0.8 M$_{\odot}$ for consistency and easier comparison with
the white dwarf-only model results. For M$_{wd} = 0.8 M_{\odot}$, and $i =
60$ degrees, the best-fitting accretion disk models had a mass accretion
rate in the range $\dot{M} = 10^{-9.5}-10^{-9} M_{\odot}$/yr, leading to a
distance of $d \approx 400-800$pc respectively. Again the fit was very
poor and the distance was not in agreement with our adopted value of
186pc.

Next, we tried combination fits of a white dwarf and accretion disk.
We follow the same procedure as described in the previous paragraph,
 namely we fix the
WD mass to $0.8 M_{\odot}$ and the inclination to 
60 degrees and we vary the mass accretion rate. 
We then varied the WD temperature for each given mass
accretion rate until we achieved a best-fit. The best fit of all 
was achieved when
the WD temperature was T$_{wd} = 27,000$K and the accretion rate was
$10^{-9.5} M_{\odot}$/yr. The $\chi^{2}_{\nu}$ 
for this fit was 3.41.  The white
dwarf contributes only 25\% of the flux and the disk contributes 75\%  
(see figure 7). However, the distance obtained was far too large 
(d$\sim$460 pc) 
and the mass accretion rate rather large for quiescence. Therefore,
again, we rejected this result. As can be seen from the figure, the
fit is better in the {\it{IUE}} range than in the {\it{FUSE}}
range, where the model flux is basically 20 \% too low.  

Finally, we compared the results of the above fitting attempts with a
two-temperature WD model fit. A WD is expected to show a temperature
variation with latitude if accretion occurs preferentially at the equator
from a disk or if a WD is magnetic and accretes preferentially at magnetic
poles. In the former case, we envision the possibility of a hot accretion
belt, or hot inner disk ring. 
We ran a series of models in which a WD is cooler and rotates more
slowly at higher latitudes and has a fast spinning (near Keplerian speed)
hot atmosphere belt. We tried a range of combinations of cooler WD and
belts of different temperatures and lower gravity. 
For the WD we kept $\log{g}=8.3$ constant and searched for a best
fit consistent with a distance of $d \approx 186$pc. The white dwarf plus
accretion belt combination which yielded the best fit has a WD with
T$_{wd} = 25,000$K, and an accretion belt with T$_{belt} = 40,000$K, 
$\log{g} = 6$ with solar abundances. 
The cooler portion of the WD contributes 75\%
of the FUV flux while the accretion belt contributes 25\% of the FUV flux.
This is the same best fit combined model as the one for the {\it{FUSE}}
spectrum alone.
The $\chi^{2}_{\nu}$ value was 3.56 and the scale parameter led to a 
distance of 190pc. 
Our comparison of the two-temperature (WD + accretion belt) 
model is displayed in figure 8. 
The improvement of model fits that
result from two-temperature WDs has been reported elsewhere for a number
of other systems (e.g. Szkody et al. 2003; Sion et al. 2003). 

\section{Conclusion}

During the quiescence of WW Ceti, our fits to the combined
{\it{FUSE}} + {\it{IUE}} fluxes reveal that a single-temperature with
$T_{wd} \sim 26,000 \pm 1000$K can account for the flux. The white
dwarf appears to have a rotational velocity of $600 \pm 100$ km$~$s$^{-1}$. 
The error bars are chosen here to be the size of the increments by which the
parameters are varied. For WW Cet, which is actually a rather weak
source for {\it{FUSE}} and {\it{IUE}}, these error bars are 
consistent with the modelings, i.e. smaller error bars/increments 
do not lead to a significant improvement in the fit.  
The best agreement with the observations is
provided by a two-temperature white dwarf model with a cooler white dwarf
at $T_{wd} = 25,000$K providing 75\% of the FUV flux and a hotter region
(accretion belt or optically thick disk ring) with T = 40,000K
contributing 25\% of the flux.  The fitting of the shape of the absorption
lines in the {\it{FUSE}} range led to the following best fit abundances: 
Carbon 0.1 x solar, Nitrogen 2 x solar, and Silicon 0.3-0.5 x solar. In all
the fitting models of the combined ({\it{FUSE}}+{\it{IUE}}) data
we have kept the mass of the WD constant ($M=0.8M_{\odot}$ corresponding
to $\log{g}=8.3$) and rejected models in disagreement with a distance of 
$d\approx 186$pc.  

An important point is the significance of the
two-temperature WD fit over the single WD fit.  A change in
chi-squared from 1.58 to 1.38 (for the {\it{FUSE}} fits) and, even more
so, from 3.73 to 3.56 (for the {\it{FUSE+IUE}} fits) is at best a modest
improvement in the fit quality with the addition of the second
temperature component. However,  
adding the second component results in a better
fit to the bottom of the Ly$\beta$ line around 1025 \AA, 
and improves the fit to
the left wing of Ly$\beta $ (the right wing is contaminated by 
the OVI emission feature). 
In that region the improvement of the fit 
is the actual fitting of the blue wing
of the Ly$\beta$ in the spectrum.  

We remark here that there could be some instrument background
contamination contributing to the flux, such that the Ly$\beta$
would actually never go to zero. However, since we have discarded
all the noisy portions of the channels (usually the edges), 
the actual contribution of the instrument contamination should be
less than $\approx 5 \times 10^{-15}$erg$~$s$^{-1}$cm$^{-2}$\AA$^{-1}$
($<$50\% of the flux) which is the excess emission at 910\AA 
(this is an overestimate, since the region $\lambda < 910$\AA\ 
is near the edge where the noise is maximal). 

In the wavelength range $\lambda <$970\AA\  the improvement of
the fit does not fit any actual feature but only reduces the
discrepancy between the model and the observation. Therefore the
need and importance of the second component does not originate from
fitting that part of the spectrum where there is emission but 
rather it comes from fitting a feature in the continuum.  
At this stage the ``belt'' is really a flat continuum added
to improve the fit and it should be regarded as a featureless
blue spectrum. So far the belt is probably not the best physical
description of the data but it is the best available model component
to help improve the fit. 
In fact, the two-temperature WD fit does not
provide the lowest $\chi^2_{\nu}$, the disk+WD does.
However, this lowest disk+WD model,
while fitting better in the {\it{IUE}} (lower resolution) spectral range,
does not provide a good fit a in the {\it{FUSE}} (higher
resolution) range of the combined spectrum and is inconsistant
with the distance of the system. 
As we stated previously, we do not chose blindly the
lowest $\chi^2_{\nu}$ model, but we chose one of the lowest $\chi^2_{\nu}$ 
models that provides a better fit to some specific parts
and features of the spectrum.  The fact that both the WD+belt
and the WD+disk provide the lowest $\chi^2_{\nu}$ models, 
reflects the fact that the second component cannot, presently, be modeled
accurately.

There are broad emission lines which are probably 
due to a hot gas, and the
O{\small{VI}} red-shifted feature may indicate the 
possibility that the material is flowing away from the observer.
However, the other emission features are not resolved enough
to confirm or refute such a scenario.     
We do not discuss here the origin or the possible
scenarios of such a flow, though the mechanisms at work
could be as varied as the ones discussed in \citet{hoa03} 
on the FUV observation of the complex system DW UMa. 
It might be worth noting that the supra-solar Nitrogen 
abundance, albeit uncertain (as it could well be from interstellar
origin), together with the sub-solar Carbon abundance, could be
a result of CNO-processing, either from a past nova or from 
CNO processed material being transferred from the processed core
of the secondary. 

Finally, our fitting results provide an additional badly needed
temperature data point in the distribution of WD $T_{wd}$ versus orbital
period above the CV period gap where few WD $T_{wd}$s are available, and
an additional white dwarf rotational velocity derived from the 
{\it{FUSE}} data
alone (see Table 3). 
The surface temperature and rotational velocity of the white dwarf 
in WW Ceti are within the range that has been derived for the handful 
of other dwarf nova white dwarfs above the period gap. The effective 
temperature, $T_{wd}=25,000$K $\pm 1000$K, 
of the white dwarf lies at the  
lower boundary of the temperature distribution for dwarf nova white 
dwarfs above the period gap. For example, at the high temperature end 
are RU Peg ($T_{wd}$ = 53,000K, \citep{sio02}) and Z Cam ($T_{wd}$ = 
57,000K; \citep{har05}) while SS Aur ($T_{wd}$ = 27,000K; \citep{sio04a}), 
U Gem ($T_{wd}$ = 30,000K; \citep{sio98}) and RX And ($T_{wd}$ 
= 34,000K; \citep{sio01,sep02}) make up the remainder of the distribution.
The rotation rate of $600 \pm 100$km$~$s$^{-1}$ is at the high end of the 
rotation rate distribution for CV white dwarfs above and below the gap 
\citep{szk02}. 

\acknowledgments This work was supported by NSF grant AST99-09155, NASA ADP
grant NNG04GE78G, NASA grant NAG5-13765, and by a
summer research assistantship from the Delaware Space Grant Consortium.
PG wishes to thank the Space Telescope Science 
Institute for its kind hospitality.

\clearpage

\begin{center}
\begin{tabular}{lc}
\multicolumn{2}{c}{Table 1:
WW Ceti System Parameters}\\ \hline\hline
Subtype:               & U Gem            \\
P$_{orb}$(d):          & 0.1758	          \\
inclination $i$ (deg): & $54\pm4$         \\
Spectral Type:         & M2.5V	          \\
M$_{1}(M_{\odot}$):    & $0.85 \pm 0.11$  \\
M$_{2}(M_{\odot}$):    & $0.41\pm 0.10$	  \\
V$_{max}$:             & 9.3	          \\
V$_{min}$:             & 13.9             \\
\hline
\end{tabular}
\end{center}
\begin{center}	
\small References: \citet{haw90,rin96,tap97,rit03}  
\end{center}

\clearpage

\begin{center}
\begin{tabular}{lcccccrc}
\multicolumn{7}{c}
{Table 2: {\it{IUE}} Observing Log}\\ \hline\hline
SWP    &  Aperture & Disp.& Date       &  Time of Observation & t$_{\rm exp}$(s) & Continuum &     Bckgr. \\ \hline
24866  &Large      & Low  & 1985 Jan 8 & 16:29:20             &	9899             &  76 cts/s &    34 cts/s  \\
\hline
\end{tabular}
\end{center}

\clearpage 

\begin{center}
\begin{tabular}{llccccl}
\multicolumn{7}{c}{Table 3:
Dward Novae Above the Gap with Known quiescent WD's Temperature}\\ \hline\hline
System     & Subtype  & $M_{wd}$      & Period  & $T_{wd}$  & $V\sin{i}$ & References \\
           &          & $(M_{\odot})$ & (hours) & (K)        & $(km~s^{-1}$)    &           \\ 
\hline 
RU Peg     &  U Gem   & 1.10          &   8.99  &  53,000    & 100   & \small{\citet{sio02}; \citet{sio04a}} \\  
Z Cam      &  Z Cam   & 0.99          &   6.96  &  57,000    & 300   & \small{\citet{har05}} \\  
EY Cyg     &  U Gem   & 1.20          &   5.24  &  22,000    & 600   & \small{\citet{sio04b}} \\    
RX And     &  Z Cam   & 1.14          &   5.04  &  34,000    & 600   & \small{\citet{sio01}; \citet{sep02}} \\   
SS Aur     &  U Gem   & 1.08          &   4.38  &  27,000    & 400   & \small{\citet{sio04a}} \\  
U Gem      &  U Gem   & 1.26          &   4.25  &  30,000    & 100   & \small{\citet{sio98}; \citet{lon99}} \\  
           &          &               &         &            &       & \small{\citet{fro01}}\\ 
WW Cet     &  U Gem?  & 0.85          &   4.22  &  25,000    & 600   & \small{This work} \\  
\hline 
\end{tabular}
\end{center}

\clearpage 

FIGURE CAPTIONS

Fig. 1
- The {\it{FUSE}} spectrum of WW Ceti.
The wavelength is given in \AA , the flux is given
in erg$~$s$^{-1}$cm$^{-2}$\AA $^{-1}$. 
We have marked the emission
and absorption lines and features. The airglow emission (sharp 
emission lines) have been annotated with a plus sign inside a circle. 
The sharp N{\small{I}} and N{\small{II}} absorption lines are probably
not associated with WW Ceti.  Note the very broad and prominent 
O{\small{VI}} doublet emission.  

Fig. 2 
- Fitting of the oxygen doublets with two Gaussians. 
Upper panel: the Gaussians have been shifted by $+1.2$\AA\ 
and are located at 
$\lambda_O$=1033.112 \AA\ \& 1038.813 \AA\ .
The Gaussians have the form 
$\propto exp - \{ \frac{\lambda-\lambda_O}{w} \} ^2  $
where $w=3.2$ \AA\ .  
Lower panel: the Gaussians are at the rest frame wavelengths 
of the Oxygen doublet (1031.912 \AA\ \& 1037.613 \AA\ ) 
and their width $w$ has been increased to $w=3.5$ \AA\ to
improve the fit.    
 
Fig. 3
- Best-fitting single temperature WD fit to the {\it{FUSE}}
spectrum of WW Ceti in quiescence: T=27,000K,
$\log{g}=8.3$ and $V\sin{i}=600$km$~$s$^{-1}$.   
The regions of the {\it{FUSE}} spectrum that have been masked are
plotted in blue. The solar abundances best-fit is in green 
and the non-solar abundances best fit is in red. The main absorption
features which are improved by the decrease of C and Si
abundances are shown with arrows. 

Fig. 4
- Residual emission in the {\it{FUSE}} spectrum after the 
non-solar abundances best fit WD model 
(presented in Figure 3) has been subtracted.  

Fig. 5
- Two-temperature (white dwarf plus accretion belt) fit to the
{\it{FUSE}} spectrum of WW Ceti in quiescence. The dotted line
represents the WD model, the dashed line represents the 
accretion belt, and the solid line represents the combined model. 
For the WD: $T_{wd} = 25,000$K, $\log{g}=8.3$ 
and $V\sin{i}=600$km$~$s$^{-1}$.   
For the accretion belt: 
$T_{belt} = 40,000$K, $\log{g}= 6$ and $V\sin{i}=3000$km$~$s$^{-1}$.   
The cooler portion
of the WD contributes 75\% of the FUV flux while the accretion belt
contributes 25\% of the FUV flux with a fractional area of only 3\%.

Fig. 6
- Best-fitting single temperature WD fit to the combined {\it{FUSE}}
plus {\it{IUE}} spectrum of WW Ceti is quiescence: 
$T_{wd}=26,000$K with $\log{g}=8.3$.

Fig. 7
- Best-fitting white dwarf plus accretion disk model to the
combined {\it{FUSE}} 
plus {\it{IUE}} spectrum of WW Ceti in quiescence.
The WD has a temperature of 27,000K and the mass accretion rate
is $\dot{M}=10^{-9.5} M_{\odot}$yr$^{-1}$. While the fit is rather
good, the distance obtained is too large by a factor of two and the
mass accretion rate is quite large for quiescence.  

Fig. 8
- Two-temperature (white dwarf plus accretion belt) fit to the
combined {\it{FUSE}} + {\it{IUE}} spectra of WW Ceti in quiescence. the
best fit has a WD with T$_{wd} = 25,000$K and an accretion belt with
T$_{belt} = 40,000$K and $\log{g}= 6$.
The cooler portion
of the WD contributes 75\% of the FUV flux while the accretion belt,
with only 3\% of the total emitting area, 
contributes 25\% of the FUV flux.
 
\clearpage 

\begin{figure}
\plotone{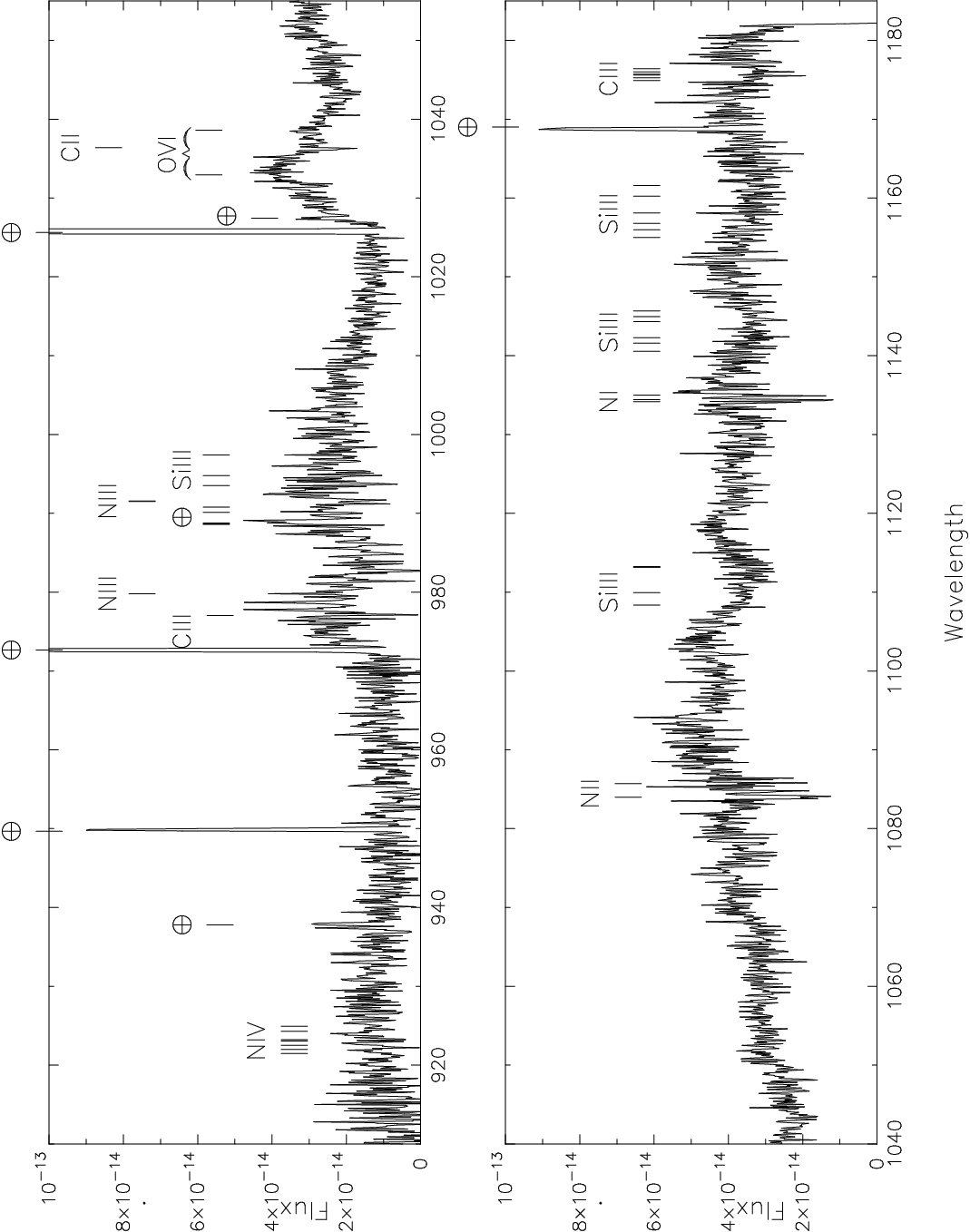} 
\end{figure} 

\clearpage 

\begin{figure}
\plotone{f2.eps} 
\end{figure} 

\clearpage 

\begin{figure}
\plotone{f3.eps} 
\end{figure} 

\clearpage 

\begin{figure}
\plotone{f4.eps} 
\end{figure} 

\clearpage 

\begin{figure}
\plotone{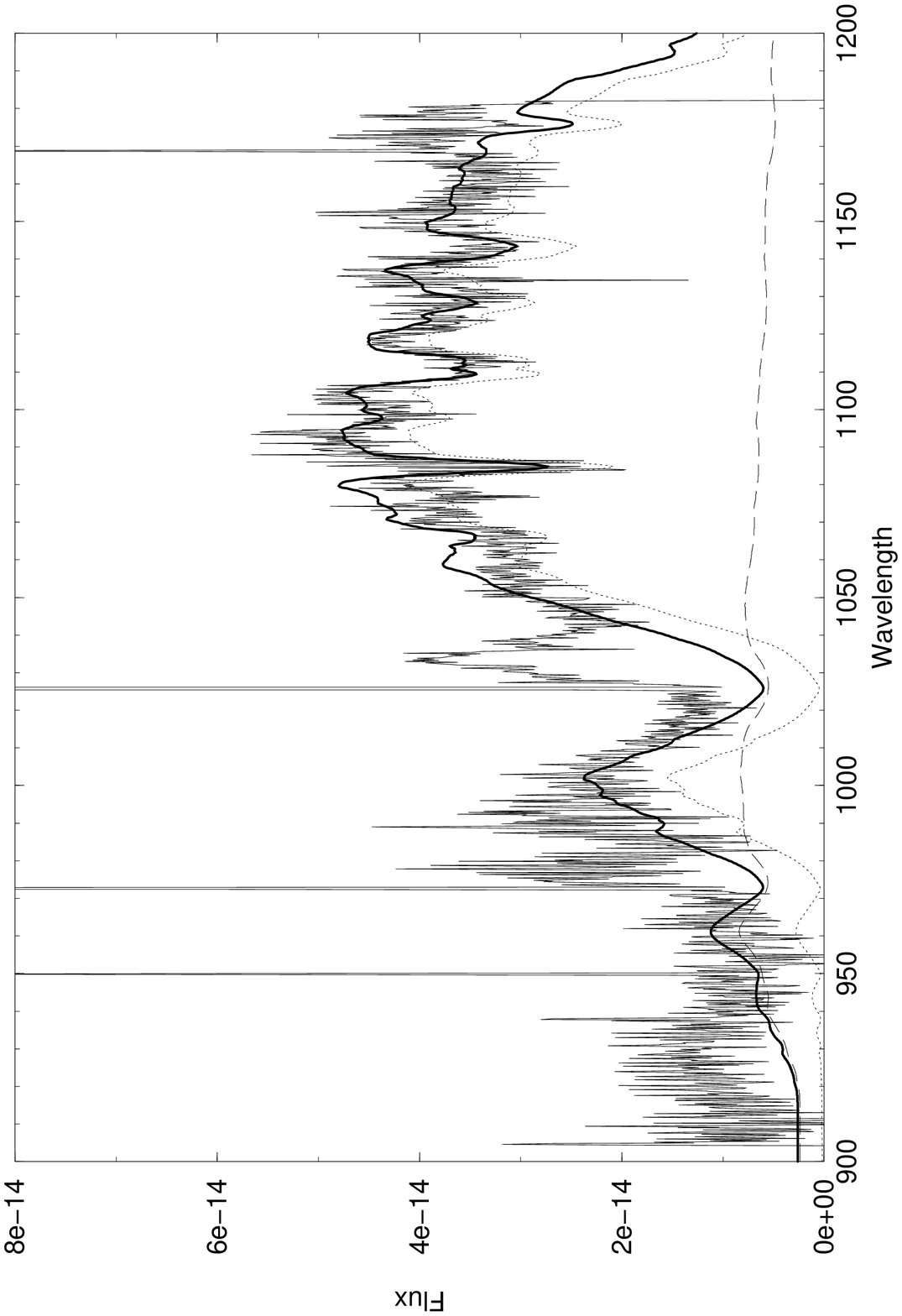}
\end{figure} 

\clearpage 

\begin{figure}
\plotone{f6.eps} 
\end{figure} 

\clearpage 

\begin{figure}
\plotone{f7.eps} 
\end{figure} 

\clearpage 

\begin{figure}
\plotone{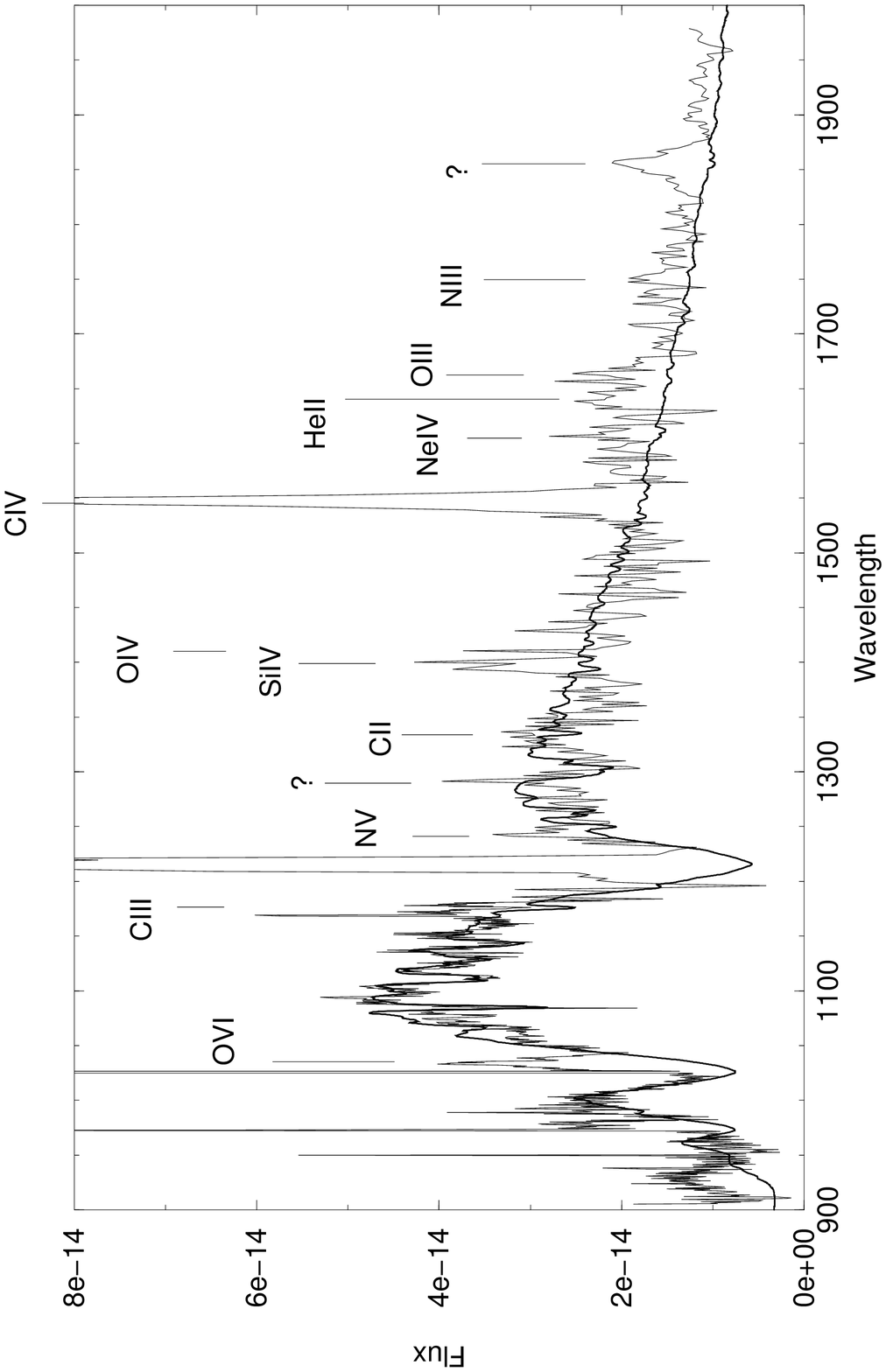} 
\end{figure} 

\end{document}